\documentclass[journal]{IEEEtran}
\usepackage{cite}      
\usepackage{multirow}  
\usepackage{booktabs}
\usepackage{makecell}
\usepackage{graphicx} 
\usepackage{soul}      
\usepackage{color, xcolor} 
\definecolor{mycolor}{RGB}{7,133,230}  
\definecolor{mycolor1}{RGB}{0,176,80}  
\definecolor{mycolor2}{RGB}{192,0,0}  
\definecolor{mycolor3}{RGB}{45,84,160} 
\usepackage{amsmath}   
\usepackage{amsfonts} 
\usepackage{algorithm}  
\usepackage{algpseudocode} 
\usepackage{cases} 
\usepackage{amssymb} 
\usepackage{amsthm}
\usepackage{bm}
\usepackage{enumitem}
\usepackage{titlesec}
\usepackage{utfsym}
\titlespacing*{\subsubsection}{0pt}{\baselineskip}{0pt}

\usepackage{threeparttable}  
\usepackage{hyperref} 
\hypersetup{ 
	colorlinks=true,
	linkcolor=blue,
	filecolor=blue,      
	urlcolor=blue,
	citecolor=blue,
}
\ifCLASSOPTIONcompsoc
\usepackage[caption=false,font=normalsize,labelfont=sf,textfont=sf]{subfig}
\else
\usepackage[caption=false,font=footnotesize]{subfig}
\fi

\ifCLASSINFOpdf
\else
\fi

\begin{document}
\title{Compound Gaussian Radar Clutter Model With \\Positive Tempered Alpha-Stable Texture} 
 
\author{Xingxing~Liao,~ 
	    Junhao~Xie,~\IEEEmembership{Senior Member,~IEEE},~Jie Zhou     

\thanks{Manuscript received XXXXX, 2024; revised XXXXX, 2024. This work was supported by the National Natural Science Foundation of China under Grant 62371162.} 
\thanks{{The authors are with the Key Laboratory of Marine Environmental Monitoring and Information Processing}, Harbin Institute of Technology, Harbin 150001, China. E-mail: (22b305010@stu.hit.edu.cn;
	xj@hit.edu.cn; 22b905037@stu.hit.edu.cn). \textit{(Corresponding author: Junhao Xie.)}}} 


\maketitle
 \begin{abstract} 
 	The compound Gaussian (CG) family of distributions has achieved great success in modeling sea clutter. 
 	This work develops a flexible-tailed CG model to improve generality in clutter modeling, by introducing the positive tempered $\alpha$-stable (PT$\alpha$S) distribution to model clutter texture. 
 	The PT$\alpha$S distribution exhibits widely tunable tails by tempering the heavy tails of the positive $\alpha$-stable (P$\alpha$S) distribution, thus providing greater flexibility in texture modeling.   
 	Specifically, we first develop a bivariate isotropic CG-PT$\alpha$S complex clutter model that is defined by an explicit characteristic function, based on which the corresponding amplitude model is derived. 
 	Then, we prove that the amplitude model can be expressed as a scale mixture of Rayleighs, just as the successful compound K and Pareto models. 
 	Furthermore, a characteristic function-based method is developed to estimate the parameters of the amplitude model. 
 	Finally, real-world sea clutter data analysis indicates the amplitude model's flexibility in modeling clutter data with various tail behaviors.   	
 \end{abstract}

 \begin{IEEEkeywords} 
 	Radar clutter, compound Gaussian (CG) model, positive tempered $\alpha$-stable (PT$\alpha$S) texture, flexible tail, parameter estimation.  
 \end{IEEEkeywords} 
 \IEEEpeerreviewmaketitle
 
 \section{Introduction}
 \label{sec:intro}  
 \IEEEPARstart{O}{ver} the past decades, significant progress has been made in the development of radar clutter models \cite{Rosenberg2019-Models,2022-Rosenberg-seaclutter}. These advancements have been crucial for developing and testing  maritime radar systems, as well as for designing target detection and tracking algorithms.  
 The compound Gaussian (CG) family has often been found as the most suitable for modeling real-world sea clutter \cite{Watts-2022-challenges}.  
 This is primarily because the CG models align with the physical formation mechanisms of sea clutter, which is described as the product of a slow-varying positive texture and the fast-varying complex Gaussian speckle \cite{1981-Ward-CG}. 
 
 The Rayleigh \cite[Chapter 6.2]{2013-Ward-seaclutter} and heavy-tailed Rayleigh (HT-Rayleigh) \cite{Kuruoglu2004-CG-Alpha} distributions are two special cases within the CG family.  
 Specifically, when radar resolution is low, invoking the central limit theorem (CLT), the in-phase (I) and quadrature (Q) components of complex clutter follow a complex Gaussian distribution, leading to the Rayleigh distribution as the amplitude distribution model.  
 As radar resolution increases, sea clutter has been observed to deviate from the Gaussian statistics. 
 One extreme case is that the clutter data show impulsive characteristics which follow the generalized CLT \cite{Ding2016-CG-Alpha,Kuruoglu2004-CG-Alpha}. 
 In this case, applying the $\alpha$-stable ($\alpha$S) law \cite[Chapter 7, \S 34]{Gnedenko-book}, the I/Q components of complex clutter follow a complex symmetric $\alpha$S (S$\alpha$S) distribution.   
 Further assuming isotropy for the I/Q components, the clutter amplitude obeys the HT-Rayleigh distribution that is a CG model with positive $\alpha$-stable (P$\alpha$S) texture \cite{Kuruoglu2004-CG-Alpha}.

 However, the amplitude distribution of real-world sea clutter often exhibits tail behaviors that are heavier than the Rayleigh model but lighter than the HT-Rayleigh model. 
 For such clutter data, various alternative CG models have shown suitability in certain cases, including K distribution \cite{Jakeman1976-K}, Pareto distribution \cite{2007-Balleri-Pareto}, and CG distribution with generalized inverse gamma texture \cite{2019Xue-CG-GIG}. 
 Nevertheless, due to the intricate physical mechanisms of sea clutter, existing CG models may not provide sufficient generality to accommodate the diverse range of data encountered in practice \cite{Watts-2022-challenges}.  
 
 The tempered $\alpha$S (T$\alpha$S) family is a generalization of the $\alpha$S family \cite{Koponen1995-TruncatedStable,Carr-2003-CTS}. 
 It tempers the heavy $\alpha$S tails to obtain widely tunable tails, so as to better align with various tail behaviors of practical data in many research fields, including in finance \cite{Miranda-2001-economics}, ecology \cite{Humphries-2012-forage}, and physics \cite{Vermeersch-2015-Conduction}.  
 The T$\alpha$S family contains two important subclasses: the symmetric T$\alpha$S (ST$\alpha$S) and positive T$\alpha$S (PT$\alpha$S) distributions, which show promise in modeling certain type of radar clutter signals.

 In \cite{Liao2024-FTRay}, recognizing that the ST$\alpha$S distribution offers tunable tail behaviors between the Gaussian and heavy-tailed S$\alpha$S distributions, we developed a bivariate isotropic ST$\alpha$S distribution to model the I/Q components of complex clutter. 
 Besides, we derive the corresponding amplitude model, named the flexible-tailed Rayleigh (FT-Rayleigh) model, which has tunable tail behaviors between the Rayleigh and HT-Rayleigh models.  
 Compared with traditional clutter models, the FT-Rayleigh model has been demonstrated to have superior performance in modeling clutter data under different sea states and polarizations. 
 However, representing the FT-Rayleigh model as a scale mixture of Rayleighs remains a challenge. 
 This limitation means that the rich theory and methods (e.g., those related to clutter simulation and target detection) developed for CG models cannot be directly applied to this model.

 To overcome this challenge, we turn to the PT$\alpha$S distribution, another important subclass of the T$\alpha$S family, to model the clutter texture. 
 This allows us to develop a new CG clutter model with flexible tails to enhance the generality of clutter modeling. 
 To be specific, we first develop a bivariate isotropic CG-PT$\alpha$S complex clutter model, defined by an explicit characteristic function. By taking the Fourier transform of this characteristic function and applying   coordinate conversions, we derive the corresponding amplitude model.   
 Next, we prove that the amplitude model is a scale mixture of Rayleighs, in accordance with other successful CG models such as the K and Pareto distributions. We therefore name it the compound FT-Rayleigh (CFT-Rayleigh). 
 Furthermore, by analogy with the method in \cite{Liao2024-FTRay}, a characteristic function-based method is developed to estimate the CFT-Rayleigh parameters. 
 Finally, analysis of real-world clutter data demonstrates that the CFT-Rayleigh model effectively characterizes sea clutter data with various tail behaviors.

 \section{Review of $\alpha$-Stable and Tempered $\alpha$-Stable Distributions} 
 \label{sec:review}  
 In the following, we review the definitions and some properties of the $\alpha$S and T$\alpha$S distributions, laying the groundwork for developing a CG clutter model with flexible tails in the next section.  
 
 \subsection{$\alpha$-Stable Distribution}  
 The characteristic function of the zero-location $\alpha$S distribution can be represented by 
 \begin{equation} 
 \begin{split} 
 & \varphi(\xi)= \exp \left[ { - \int_{ - \infty }^\infty  {\left( {1 - {e^{ - j\xi y}}} \right)} f(dy)} \right]    
 \label{eq-stable-CF-CompoundPoisson} 
 \end{split} 
 \end{equation}
 where $j$ is the imaginary unit, and $f(dy)$ is a L{\'e}vy measure defined by
 \begin{equation} 
 \begin{split} 
 f(dy) = \left(A_- \mathbf{1}_{\{y < 0\}} + A_+ \mathbf{1}_{\{y \geq 0\}}\right) |y|^{-\alpha-1} dy
 \label{eq-stable-Levy} 
 \end{split}  
 \end{equation} 
 where $A_-,A_+ >0$ are scale constants, $\alpha \in(0,2]$ is the characteristic exponent, and $\mathbf{1}_{\{\cdot\}}$ denotes the indicator function. 
 Substituting (\ref{eq-stable-Levy}) into (\ref{eq-stable-CF-CompoundPoisson}), the characteristic function of an $\alpha$S random variable (RV) is  
 \begin{equation} 
 \begin{split} 
 \varphi(\xi)= \exp \left[ {   - \gamma |\xi {|^\alpha }\left( {1 + j\beta {\mathop{\rm sign}\nolimits} (\xi )\omega (\xi ,\alpha )} \right)} \right] 
 \label{eq-stable-CF} 
 \end{split} 
 \end{equation} 
 where $\gamma=\left(A_{+}+A_{-}\right) \pi \cos (\pi \alpha / 2) /[\alpha \Gamma(\alpha) \sin (\pi \alpha)]>0$ is the scaling factor, and $\beta = (A_+ - A_-)/(A_+ + A_-) \in [-1,1]$  denotes the skewness. 
 $\operatorname{sign}(\cdot)$ is the signum function, $|\cdot|$ denotes the absolute value, and
 \begin{equation} 
 \begin{split} 
 \omega(\xi, \alpha) = \left\{\begin{array}{rr}
 \tan (\pi \alpha / 2) & \text { if } \alpha \neq 1, \\
 (2 / \pi) \ln |\xi| & \text { if } \alpha=1.  
 \end{array}\right.
 \end{split} 
 \notag
 \end{equation} 
 
 The $\alpha$S distribution defined by (\ref{eq-stable-CF}) has two important subclasses. 
 Specifically, when $\beta=0$, it simplifies to the symmetric subclass, i.e., the S$\alpha$S distribution, which further reduces to the Gaussian distribution when $\alpha=2$. 
 When $\alpha \in (0,1)$ and $\beta=-1$, the $\alpha$S distribution is totally skewed to the right along the positive real axis, reducing to the P$\alpha$S distribution.   
 
 It has been reported in various research fields that real-world data exhibit heavier tails than the Gaussian distribution but lighter tails than the $\alpha$S distribution \cite{Miranda-2001-economics,Humphries-2012-forage,Vermeersch-2015-Conduction}. 
 To capture this observed tail behavior, researchers proposed the T$\alpha$S distribution to lighten the tails of the $\alpha$S distribution, as introduced below.

 \subsection{Tempered $\alpha$-Stable Distribution}   
 Noting that the L{\'e}vy measure (\ref{eq-stable-Levy}) are intimately related to the $\alpha$S tails, Koponen \cite{Koponen1995-TruncatedStable} proposed to temper it so as to lighten the tails of the $\alpha$S distribution. 
 Specifically, the L{\'e}vy measure (\ref{eq-stable-Levy}) was tempered to  \cite{Koponen1995-TruncatedStable,Carr-2003-CTS} 
 \begin{equation} 
 \begin{split} 
 f(dy) =  \left(A_- \mathbf{1}_{\{y < 0\}} + A_+ \mathbf{1}_{\{y \geq 0\}}\right)|y|^{-\alpha-1} q(y) dy
 \label{eq-stable-Levy-Trunc} 
 \end{split}  
 \end{equation}
 where $q(y)$ is the tempering function defined by 
 \begin{equation} 
 \begin{split} 
 q(y) =  \exp(y/\eta_-) \mathbf{1}_{\{y < 0\}} + \exp(-y/\eta_+)\mathbf{1}_{\{y \geq 0\}}  
 \end{split}  
 \end{equation} 
 where $\eta_-$ and $\eta_+$ are the truncation parameters.  
 By replacing the L{\'e}vy measure in (\ref{eq-stable-CF-CompoundPoisson}) with (\ref{eq-stable-Levy-Trunc}), the characteristic function of the T$\alpha$S distribution can be obtained. 
 When $\eta_-=\eta_+=\eta$, the T$\alpha$S distribution reduces to the smoothly truncated L{\'e}vy flight  \cite{Koponen1995-TruncatedStable}. 
 In this case, the T$\alpha$S characteristic function is given by (\ref{eq-CTS-CF}), 
 \begin{figure*}[t]
 	\centering
 	\begin{equation}
 	\begin{split} 
 	\varphi (\xi ) &= \exp \left\{ {  - {A_ + }\Gamma \left( { - \alpha } \right)\left[ {{\eta ^{ - \alpha }} - {{\left( {{\eta ^{ - 1}} + j\xi } \right)}^{ - \alpha }}} \right] - {A_ - }\Gamma \left( { - \alpha } \right)\left[ {{\eta ^{ - \alpha }} - {{\left( {{\eta ^{ - 1}} - j\xi } \right)}^\alpha }} \right]} \right\}\\
 	&=\exp \left\{ {  \frac{\gamma }{{{\eta ^\alpha }\cos \frac{{\pi \alpha }}{2}}}\left[ {1 - {{\left( {{{\left( {\eta \left| \xi  \right|} \right)}^2} + 1} \right)}^{\frac{\alpha }{2}}}\cos \left( {\alpha \arctan \left( {\eta \left| \xi  \right|} \right)} \right)\left[ {1 + j{\rm{sign}}(\xi )\beta \tan \left( {\alpha \arctan \left( {\eta \left| \xi  \right|} \right)} \right)} \right]} \right]} \right\} 
 	\label{eq-CTS-CF} 
 	\end{split}  
 	\end{equation}
  	\vspace*{8pt} 
	\hrulefill
	\vspace*{8pt} 
 \end{figure*} 
 where $\gamma$ and $\beta$ have the same meanings as in the $\alpha$S distribution (\ref{eq-stable-CF}).     
 
 It is worth mentioning that by generalizing the tempering function $q(y)$, various T$\alpha$S distributions have been obtained so far, including generalized T$\alpha$S, modified T$\alpha$S, and rapidly decreasing T$\alpha$S distributions (see \cite{Grabchak2016-Trun,Fallahgoul-2021-TS,Massing-2023-TS-Estimation} for details). 
 In this work, we focus on the classical T$\alpha$S distribution defined by (\ref{eq-CTS-CF}) since it remains relatively simple in form yet provides sufficient tail flexibility.

 Let $TS_{\alpha} \left (\gamma, \beta, \eta \right )$ denote the T$\alpha$S distribution defined by (\ref{eq-CTS-CF}), which reduces to the $\alpha$S distribution when $\eta \to \infty$. 
 Similar to the $\alpha$S family, there are two important subclasses of the T$\alpha$S family, i.e., the symmetric T$\alpha$S (ST$\alpha$S) and positive T$\alpha$S (PT$\alpha$S) distributions. 
 Specifically, the ST$\alpha$S distribution is $TS_{\alpha} \left (\gamma, 0,\eta \right )$; the PT$\alpha$S distribution is $TS_{\alpha/2} \left (\gamma, -1,\eta \right )$, which concentrates on the positive real axis.

\begin{figure*}[t] 
	\vspace{-0.2cm} 
	\centering
	\includegraphics[width=0.96\linewidth]{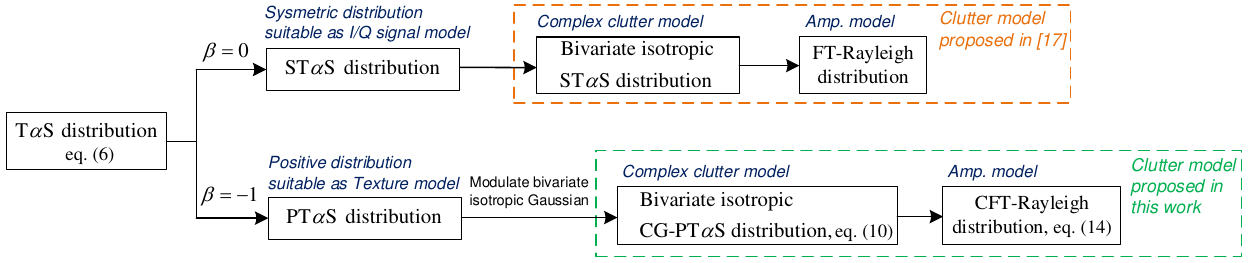} 
	\caption{Sketch of the construction of the proposed CFT-Rayleigh model and its distinction from the FT-Rayleigh model in \cite{Liao2024-FTRay}.}  
	\label{fig-explain-Flowchart} 
\end{figure*}

 When modeling sea clutter, the tail region often attracts special attention due to its critical role in detecting weak targets. 
 Therefore, developing a model that has sufficient flexibility in describing clutter tail behaviors is of great importance. 
 The above analysis suggests that ST$\alpha$S and PT$\alpha$S distributions within the T$\alpha$S family have widely tunable tail behaviors, implying their potential use in clutter modeling. 
 As shown in Fig. \ref{fig-explain-Flowchart}, the ST$\alpha$S and PT$\alpha$S distributions are suitable to model clutter I/Q components and texture, respectively, according to their distinct properties. 
 In \cite{Liao2024-FTRay}, a clutter model based on the ST$\alpha$S distribution is presented. 
 However, representing this model in a CG form remains a challenge. 
 Considering the physical formation mechanisms of sea clutter, it is typically preferable to maintain compatibility with the well-established CG family.   
 Therefore, in this work, we focus on developing a CG clutter model with PT$\alpha$S texture, which is presented in the following section.      

 \section{CG Clutter Model with PT$\alpha$S Texture}  
 \label{sec:Proposed models} 
 In this section, we first introduce a bivariate isotropic CG-PT$\alpha$S distribution to model complex clutter and derive the corresponding amplitude model. 
 Then, we prove that the amplitude model can be expressed as a scale mixture of Rayleighs. 
 Finally, we devise a characteristic function-based parameter estimation method for the amplitude model to facilitate its practical use.

 \subsection{Bivariate Isotropic CG-PT$\alpha$S Complex Signal Model and Its Resulting Amplitude Model} 
 \subsubsection{Bivariate Isotropic CG-PT$\alpha$S Complex Signal Model}  
 The bivariate CG-PT$\alpha$S complex clutter model can be represented as 
 \begin{equation} 
 \begin{split} 
 X_\text{I}+jX_\text{Q}  = \sqrt{V}(Z_\text{I}+jZ_\text{Q}) 
 \label{eq-complex-clutter-CG} 
 \end{split} 
 \end{equation} 
 where $X_\text{I}$ and $X_\text{Q}$ represent the I/Q components of the received complex signal.   
 $Z_\text{I}$ and $Z_\text{Q}$ represent the I/Q components of the complex Gaussian speckle, $Z_\text{I}, Z_\text{Q}  \sim \mathcal{N}(0,2)$. 
 $V$ denotes the PT$\alpha$S texture, $V \sim TS_{\alpha/2} \left (\cos(\pi\alpha/4)\gamma, -1,\eta \right )$. 
 $V$ and $Z_\text{I}+jZ_\text{Q}$ are independent. 
 
 It needs to be highlighted that the bivariate CG-PT$\alpha$S complex clutter model (\ref{eq-complex-clutter-CG}) can be regarded as a special case of a finance model, named multivariate normal tempered $\alpha$S \cite{Kim-2012-NTS,Bianchi-2022-NTS}. 
 This multivariate normal tempered $\alpha$S distribution is defined as 
 \begin{equation} 
 \begin{split} 
 \mathbf{X} = \mu + \theta{V}+\sqrt{V} \mathbf{Z}
 \label{eq-NTS} 
 \end{split} 
 \end{equation}
 where $\mu,\theta \in \mathbb{R}$. $\mathbf{Z}\sim \mathcal{N}(\mathbf{0},2\boldsymbol{\Sigma})$ is a $1\times d$ Gaussian random vector, with $\boldsymbol{\Sigma}$ denoting the $d\times d$ covariance matrix of $\mathbf{Z}$. $V\sim TS_{\alpha/2} \left (\cos(\pi\alpha/4)\gamma, -1,\eta  \right )$ is a PT$\alpha$S RV. $V$ and $\mathbf{Z}$ are independent. 
 
 Thus, our bivariate CG-PT$\alpha$S complex clutter model (\ref{eq-complex-clutter-CG}) can be derived from (\ref{eq-NTS}) by setting $\mu=\theta=0$, $d=2$, and $\mathbf{X}=[X_\text{I},X_\text{Q}]$ and $\mathbf{Z}=[Z_\text{I},Z_\text{Q}]$. 
 According to \cite{Bianchi-2022-NTS}, $\mathbf{X}$ can be defined by the characteristic function as follows: 
 \begin{equation} 
 \begin{split} 
 {\varphi _{\mathbf{X}}}({\boldsymbol{\xi}} ) = \exp \left( {\frac{\gamma }{{{\eta ^{\frac{\alpha }{2}}}}}\left[ {1 - {{\left( {\eta {\boldsymbol{\xi}^T}\boldsymbol{\Sigma} {\boldsymbol{\xi }} + 1} \right)}^{\frac{\alpha }{2}}}} \right]} \right) 
 \label{eq-NTS-CF-sys-zeromean-muti} 
 \end{split} 
 \end{equation} 
 where $\boldsymbol{\xi}   = \left[ {{\xi_1},{\xi_2}} \right]$ denotes the frequency vector and $(\cdot)^T$ denotes the transposition. 
 
 Moreover, given that the isotropic assumption has achieved great success in the field of modeling radar clutter, we assume that $X_\text{I}$ and $X_\text{Q}$ are isotropic, which leads to $\boldsymbol{\Sigma}$ being an identity matrix.  
 Thus, the characteristic function of the bivariate isotropic CG-PT$\alpha$S complex clutter model is  
 \begin{equation} 
 \begin{split} 
 \varphi_{\mathbf{X}}(\xi_1,\xi_2) = \exp \left( {\frac{\gamma }{{{\eta ^{\frac{\alpha }{2}}}}}\left[ {1 - {{\left( {\eta s^2 + 1} \right)}^{\frac{\alpha }{2}}}} \right]} \right) 
 \label{eq-NTS-CF-sys-zeromean-Biv} 
 \end{split} 
 \end{equation} 
 where $s=\|\boldsymbol{\xi}\| = \sqrt {\xi_1^2 + \xi_2^2} $. 

 As many practical applications rely on the amplitude distribution of complex clutter, we derive the amplitude model from (\ref{eq-NTS-CF-sys-zeromean-Biv}) in the following.

 \subsubsection{Resulting Amplitude Model} 
 By taking the Fourier transform of (\ref{eq-NTS-CF-sys-zeromean-Biv}), we get the joint distribution of $X_\text{I}$ and $X_\text{Q}$ as follow: 
 \begin{equation} 
 \begin{split} 
 &f_{\mathbf{X}}\left( {{x_{{\text{I}}}},{x_{{\text{Q}}}}} \right) = \frac{1}{4\pi^2 } \times\\
 &\int_{{\xi_1}} \int_{{\xi_2}} \varphi_{\mathbf{X}} ({\xi_1},{\xi_2}) \exp \left( { - j\left( {{x_{{\text{I}}}}{\xi_1} + {x_{{\text{Q}}}}{\xi_2}} \right)} \right)d{\xi_1}d{\xi_2}.  
 \label{eq-Biv-NTS-PDF} 
 \end{split} 
 \end{equation} 
 
 Defining $s  =\sqrt {\xi_1^2 + \xi_2^2}$ and $\theta  = \arctan \left( {{\xi _1}/{\xi _2}} \right)$, we can convert (\ref{eq-Biv-NTS-PDF}) into the polar coordinates as 
 \begin{equation} 
 \begin{split} 
 &f_{\mathbf{X}}\left( {{x_{{\text{I}}}},{x_{{\text{Q}}}}} \right) = \frac{1}{4\pi^2 } \int_0^\infty  s 
 \exp \left( {\frac{\gamma }{{{\eta ^{\frac{\alpha }{2}}}}}\left[ {1 - {{\left( {\eta s^2 + 1} \right)}^{\frac{\alpha }{2}}}} \right]} \right)  \\
 &\qquad  \times \int_0^{2\pi } {\exp \left( { - j{x_{{\text{I}}}}s\sin\theta - j{x_{{\text{Q}}}}s\cos\theta} \right)} d\theta ds. 
 \label{eq-Biv-NTS-PDF-polar} 
 \end{split} 
 \end{equation} 
 
 Further if we define $r = \sqrt {x_{{\text{I}}}^2 + x_{{\text{Q}}}^2} $ and $\phi  = \arctan \left( {{x_{{\text{Q}}}}/{x_{{\text{I}}}}} \right)$, (\ref{eq-Biv-NTS-PDF-polar}) can be converted into the amplitude-phase domain as  
 \begin{equation} 
 \begin{split} 
 &f_{{R},{\Phi}}\left( {r,\phi } \right) = r f_{\mathbf{X}}\left( {r\cos\phi,r\sin\phi} \right)\\
 & =  \frac{r}{2\pi } \int_0^\infty  s \exp \left( {\frac{\gamma }{{{\eta ^{\frac{\alpha }{2}}}}}\left[ {1 - {{\left( {\eta s^2 + 1} \right)}^{\frac{\alpha }{2}}}} \right]} \right)  {J_0}\left( {rs} \right)ds
 \label{eq-Biv-NTS-PDF-Rtheta} 
 \end{split} 
 \end{equation} 
 where ${J_0}(\cdot)$ is the zeroth-order Bessel function of the first kind.

 By integrating $\phi$ in (\ref{eq-Biv-NTS-PDF-Rtheta}) from $0$ to $2\pi$, we get the final expression of the amplitude PDF with unknown parameters $\{\alpha,\gamma,\eta\}$ as 
 \begin{equation} 
 \begin{split} 
 &f\left( r \right) =  r\int_0^\infty  s \exp \left( {\frac{\gamma }{{{\eta ^{\frac{\alpha }{2}}}}}\left[ {1 - {{\left( {\eta s^2 + 1} \right)}^{\frac{\alpha }{2}}}} \right]} \right)  {J_0}\left( {rs} \right)ds.
 \label{eq-Biv-NTS-PDF-R} 
 \end{split} 
 \end{equation} 
 When $\eta \to \infty$, (\ref{eq-Biv-NTS-PDF-R}) reduces to the HT-Rayleigh distribution, where $\alpha=2$ corresponds to the Rayleigh distribution. 
 
 The amplitude complementary cumulative distribution function (CCDF) corresponding to (\ref{eq-Biv-NTS-PDF-R}) is 
 \begin{equation} 
 \begin{split} 
 &\bar{F} \left( r \right) = 1-\\
 &\int_0^r {\tau \int_0^\infty  s \exp \left( {\frac{\gamma }{{{\eta ^{\frac{\alpha }{2}}}}}\left[ {1 - {{\left( {\eta s^2 + 1} \right)}^{\frac{\alpha }{2}}}} \right]} \right)  {J_0}\left( {\tau s} \right)dsd\tau}.  
 \label{eq-Biv-NTS-CCDF-R} 
 \end{split} 
 \end{equation}

 Although the amplitude PDF (\ref{eq-Biv-NTS-PDF-R}) is derived from the bivariate isotropic CG-PT$\alpha$S complex clutter model, its connection to the CG family is not apparent from its expression. 
 Therefore, the following subsection further demonstrates that (\ref{eq-Biv-NTS-PDF-R}) can be expressed as a scale mixture of Rayleighs, confirming that it is a CG model.

 \subsection{Amplitude Model Represented as Rayleigh Mixtures} 
 For the bivariate CG-PT$\alpha$S complex model, i.e., $\mathbf{X} = \sqrt{V} \mathbf{Z}$, defined by (\ref{eq-NTS-CF-sys-zeromean-muti}), its PDF can be expressed as a scale mixture of multivariate Gaussians:  
 \begin{equation} 
 \begin{split} 
 {f_{\mathbf{X}}}({\mathbf{x}} ) & = \int_{0}^\infty  {{f_{{\bf{X}},V}}\left( {{\bf{x}},v} \right)dv}  =\int_{0}^\infty  {{f_{{\mathbf{X}}\left| V \right.}}\left( {{\mathbf{x}}\left| v \right.} \right){f_V}( v)dv} \\ 
 & = \int_0^\infty  {\frac{1}{{{{  {4\pi v} }}{\sqrt{\det (\boldsymbol{\Sigma})}}}}} \exp \left( { - \frac{{{{\mathbf{x}}^T}{\boldsymbol{\Sigma}^{ - 1}}{\mathbf{x}}}}{{4v}}} \right){f_V}(v)dv 
 \label{eq-CTS-MixPDF-multi} 
 \end{split} 
 \end{equation}
 where ${f_V}$ is the mixing distribution, $V \sim TS_{\alpha/2} \left (\cos(\pi\alpha/4)\gamma, -1,\eta \right )$. ${f_{\mathbf{X}|V}}$ is the conditional distribution of $\mathbf{X}$ given $V$. 
 Specifically, for a fixed $V=v$, we have $ \mathbf{X}|V = v^{1/2}\mathbf{Z} \sim \mathcal{N}(\mathbf{0},2v\boldsymbol{\Sigma})$. $\det(\cdot)$ denotes the determinant of a matrix. 
 
 Further, under the isotropic assumption, $\boldsymbol{\Sigma}$ is an identity matrix. 
 Thus, the PDF of the bivariate isotropic CG-PT$\alpha$S distribution is obtained as follows:
 \begin{equation} 
 \begin{split} 
 {f_{\mathbf{X}}}(\mathbf{x} ) = \int_0^\infty  {\frac{1} {4\pi v}} \exp \left( { - \frac{r^2}{{4v}}} \right){f_V}(v)dv. 
 \label{eq-CTS-MixPDF-bi} 
 \end{split} 
 \end{equation}

 The corresponding amplitude PDF is calculated as 
 \begin{equation} 
 \begin{split} 
 {f}(r) &= \int_0^{2\pi } {r{f_\mathbf{X}}( \mathbf{x} )} d\phi =\int_0^\infty  {\frac{r} {2v}} \exp \left( { - \frac{r^2}{{4v}}} \right){f_V}(v)dv.  
 \label{eq-CTS-MixPDF-bi-amp} 
 \end{split} 
 \end{equation}

 The amplitude PDFs (\ref{eq-Biv-NTS-PDF-R}) and (\ref{eq-CTS-MixPDF-bi-amp}) are both derived from the bivariate isotropic CG-PT$\alpha$S complex model, but from different perspectives. Next, we provide a direct proof that (\ref{eq-Biv-NTS-PDF-R}) equals (\ref{eq-CTS-MixPDF-bi-amp}).

 \textit{Proof}: 
 According to \cite{Ridout2009-Laplace}, the laplace transform of the PDF of the PT$\alpha$S RV, i.e., $V \sim TS_{\alpha/2} \left (\cos(\pi\alpha/4)\gamma, -1,\eta \right )$, is 
 \begin{equation} 
 \begin{split} 
 {f_V^*}({s^\prime }) &= \int_0^\infty  {\exp \left( { - s^\prime v} \right)} {f_V}\left( v \right)dv \\
 &=\exp \left[ {\frac{\gamma }{{{\eta ^{\alpha /2}}}}\left( {1 - {{\left( {\eta s^\prime + 1} \right)}^{\alpha /2}}} \right)} \right].  
 \label{eq-PCTaS-laplace} 
 \end{split} 
 \end{equation} 
 
 Therefore, we can rewrite (\ref{eq-Biv-NTS-PDF-R}) as follows:  
 \begin{equation} 
 \begin{split} 
 f\left( r \right) &=  r\int_0^\infty  s \exp \left( {\frac{\gamma }{{{\eta ^{{\alpha}/{2}}}}}\left[ {1 - {{\left( {\eta s^2 + 1} \right)}^{{\alpha}/{2}}}} \right]} \right)  {J_0}\left( {rs} \right)ds\\
 & = r\int_0^\infty  s {f_V^*}({s^2}){J_0}\left( {rs} \right)ds \\
 & = r\int_0^\infty  {\int_0^\infty  {s\exp \left( { - {s^2}v} \right)} {J_0}\left( {rs} \right)ds} {f_V}\left( v \right)dv. 
 \label{eq-Biv-NTS-PDF-R-proof} 
 \end{split} 
 \end{equation} 

 By applying \cite[eq. (7.421.2)]{Gradshteyn2014-book}, the inner integral in (\ref{eq-Biv-NTS-PDF-R-proof}) can be easily solved as follows: 
 \begin{equation} 
 \begin{split} 
 {\int_0^\infty  {s\exp \left( { - {s^2}v} \right)} {J_0}\left( {rs} \right)ds} = \frac{1}{{2v}}\exp \left( {\frac{{{r^2}}}{{4v}}} \right). 
 \label{eq-Biv-NTS-PDF-R-proof2}  
 \end{split} 
 \end{equation} 
 
 Substituting (\ref{eq-Biv-NTS-PDF-R-proof2}) into (\ref{eq-Biv-NTS-PDF-R-proof}), we arrive at (\ref{eq-CTS-MixPDF-bi-amp}), which completes the proof. $\hfill\square$

\begin{figure}[t] 
 	\vspace{-0.4cm}
 	\centering
 	\subfloat[]{ 
 		\includegraphics[width=0.48\linewidth]{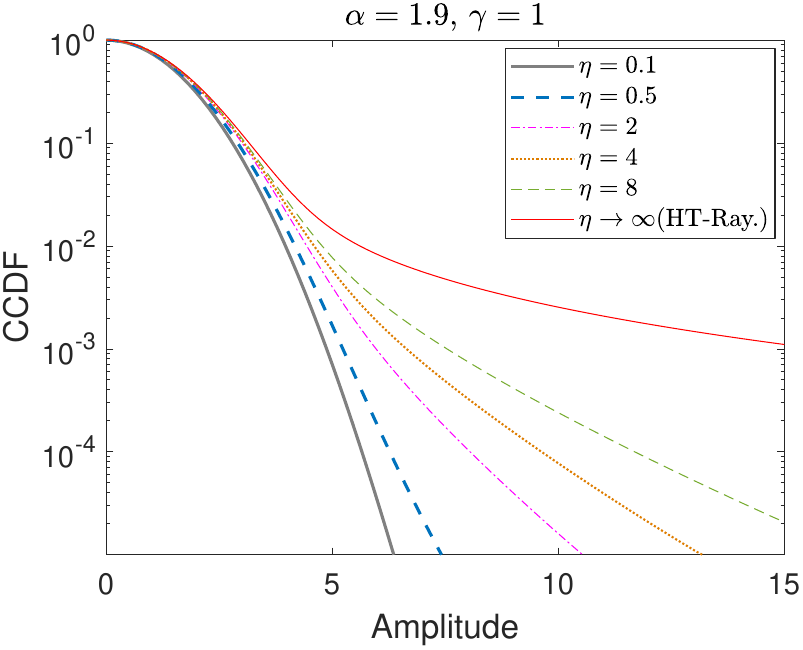} 
 	} 
 	\subfloat[]{ 
 		\includegraphics[width=0.48\linewidth]{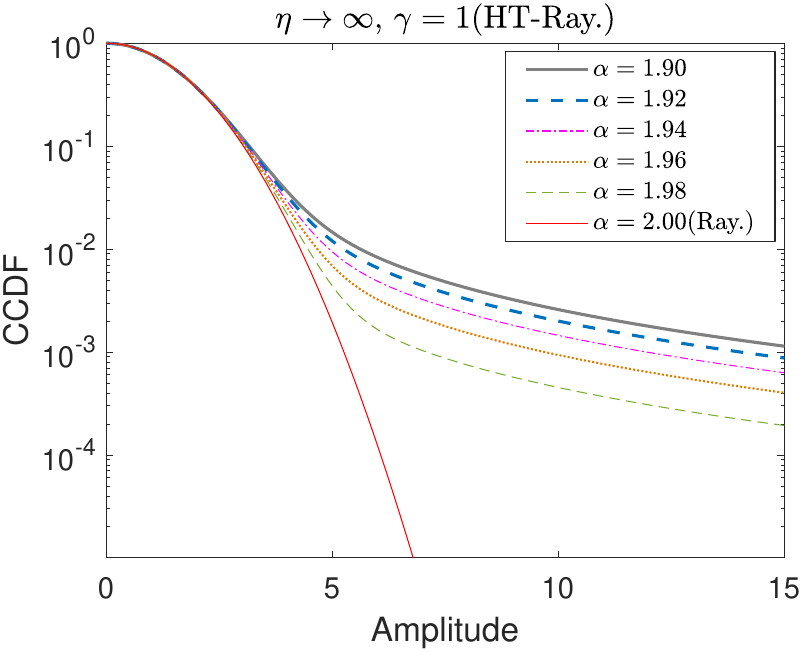}
 	}	
 	\caption{CFT-Rayleigh CCDFs with varying parameters. (a) $\alpha=1.9$, $\gamma=1$ and varying $\eta$. (b) $\eta \to \infty$, $\gamma=1$ and varying $\alpha$ (reducing to the HT-Rayleigh special case).}  
 	\label{fig-HTRay-CFTRay} 
 \end{figure}

 The above analysis motivates us to name the amplitude distribution (\ref{eq-Biv-NTS-PDF-R}) CFT-Rayleigh, so as to highlight its nature as a Compound model with Flexible Tail behavior ranging between the Rayleigh and HT-Rayleigh distributions.  
 Fig. \ref{fig-HTRay-CFTRay}(a) presents the CCDFs of the CFT-Rayleigh distribution with $\alpha=1.9$, $\gamma=1$, and varying $\eta$; Fig. \ref{fig-HTRay-CFTRay}(b) shows the HT-Rayleigh special case, characterized by $\eta \to \infty$, $\gamma=1$, and varying $\alpha$. 
 For the HT-Rayleigh CCDFs, it is evident that the tail remains relatively light only when $\alpha=2$ (Rayleigh). However, as $\alpha$ decreases below 2, the tail transitions sharply to a much heavier profile. 
 On the other hand, the CFT-Rayleigh CCDFs demonstrate that adjusting $\eta$ provides much greater flexibility in tail behavior.

 \begin{figure*}[t] 
 	\centering
 	\subfloat[]{
 		\includegraphics[width=0.325\linewidth]{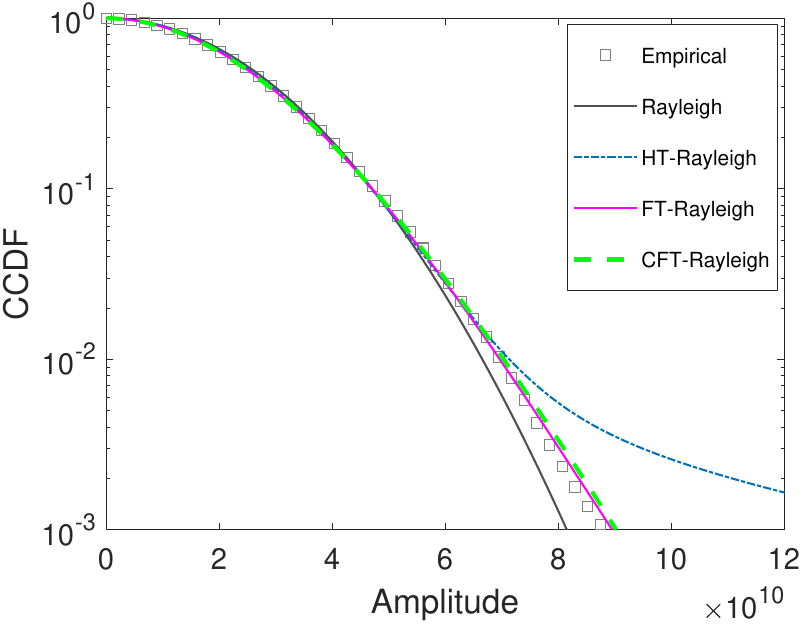}   
 	}	
 	\subfloat[]{ 
 		\includegraphics[width=0.32\linewidth]{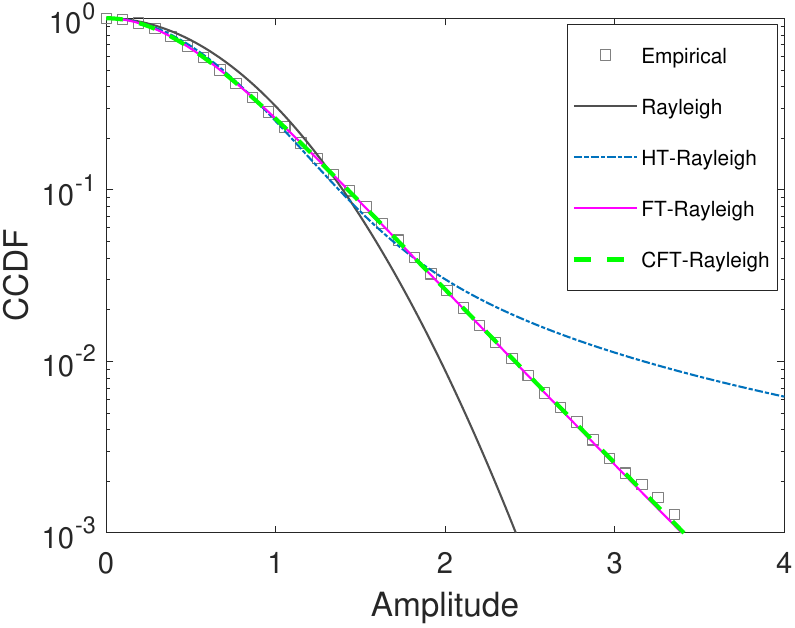}    
 	}	
 	\subfloat[]{ 
 		\includegraphics[width=0.32\linewidth]{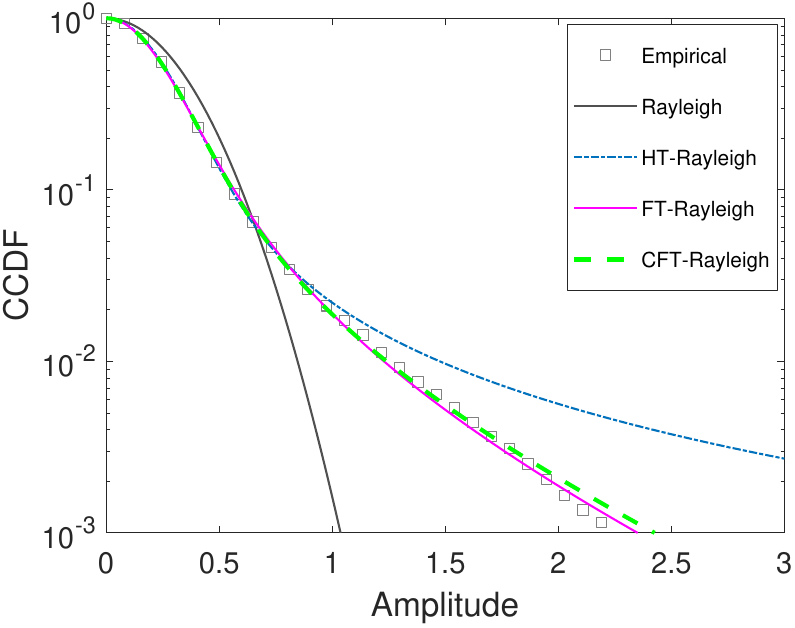}
 	}	
 	\caption{CCDF fit for CSIR Fynmeet radar Data\#1-\#3 of VV polarization. (a) Data\#1, range gate 7. (b) Data\#2, range gate 36. (c) Data\#3, range gate 9.}  
 	\label{fig-Data-CSIR-AMP-fit} 
 \end{figure*}  
 
 \subsection{Characteristic Function-based Estimator} 
 To apply the CFT-Rayleigh distribution in practice, we must estimate its parameters based on real-world data.
 Given the CFT-Rayleigh distribution defined by (\ref{eq-NTS-CF-sys-zeromean-Biv}), we estimate its parameters by matching the theoretical characteristic function (\ref{eq-NTS-CF-sys-zeromean-Biv}) to its empirical counterpart. 
 The estimation process follows the approach outlined in \cite{Liao2024-FTRay}. 
 Here, we briefly introduce the main steps.

 First, to mitigate the complexity of the subsequent estimation, we construct a theoretical normalized characteristic function (NCF) to temporally eliminate the scale parameter $\gamma$. That is 
 \begin{equation} 
 \begin{split} 
 u(s; \alpha,\eta) &= \frac{\ln \varphi_\mathbf{X}(\xi_1,\xi_2)}{\ln \varphi_\mathbf{X}(\xi_\text{ref},\xi_\text{ref})} =  \frac{  1 - {\left( {\eta {s^2} + 1} \right)^{\frac{\alpha }{2}}}   }{ 1 - {\left( {\eta {s_\text{ref}^2} + 1} \right)^{\frac{\alpha }{2}}} } 
 \label{eq-STSaS-CF-ln-ratio} 
 \end{split} 
 \end{equation} 
 where $s_\text{ref} = \sqrt{2}\xi_\text{ref}$ is the reference frequency point, set to $s_\text{ref} = 1/{P}$ with $P$ denoting the square root of the power of the amplitude samples. 
 Let $\left\{r_{i};i = 1,\cdots, L\right\}$ denote $L$ amplitude samples, we have $P=(\frac{1}{L} \textstyle \sum_{i=1}^{L}r_i^2)^{1/2}$. 
 Note that as the power of an amplitude sequence increases, the PDF elongates along the amplitude axis, leading to a compression of the characteristic function along the frequency axis. 
 Thus, division by $P$, here and below in (\ref{eq-STSaS-CF-ln-ratio-LS}), serves to eliminate the power impact and confine the frequency points within the effective zone. 
 
 Then, we need to construct the empirical NCF corresponding to (\ref{eq-STSaS-CF-ln-ratio}). 
 For a bivariate isotropic distribution, the empirical characteristic function can be expressed as a function of the amplitude samples, i.e., $\hat \varphi_\mathbf{X} \left( {{\xi _1},{\xi _2}} \right) = \frac{1}{L}  {\textstyle \sum_{i=1}^{L}}  {{J_0}\left( {s{r_i}} \right)}$.  
 Therefore, the empirical NCF is constructed as 
 \begin{equation} 
 \begin{split} 
 \hat{u}(s) &= \frac{\ln \hat\varphi_\mathbf{X}(\xi_1,\xi_2)}{\ln \hat\varphi_\mathbf{X}(\xi_\text{ref},\xi_\text{ref})} 
 = \frac{{\ln \left[ {\frac{1}{L}\sum\nolimits_{i = 1}^L {{J_0}\left( {s{r_i}} \right)} } \right]}}{{\ln \left[ {\frac{1}{L}\sum\nolimits_{i = 1}^L {{J_0}\left( {{s_{{\rm{ref}}}}{r_i}} \right)} } \right]}}. 
 \label{eq-STSaS-CF-ln-ratio-empi} 
 \end{split} 
 \end{equation}  
 
 Let $\left\{s_{k}; k = 1,\cdots, K\right\}$ denote the positive zero points of the Hermitian polynomials of order $2K$, and $\left\{w_{k}; k = 1,\cdots, K\right\}$ denote the corresponding weights \cite[Table 25.10]{Abramowitz1968handbook}. 
 Then, the characteristic exponent $\alpha$ and truncation parameter $\eta$ can be estimated by applying the Gauss-Hermite quadrature to match (\ref{eq-STSaS-CF-ln-ratio}) and (\ref{eq-STSaS-CF-ln-ratio-empi}) as 
 \begin{equation} 
 \begin{split} 
 \hat \alpha, \hat \eta  = \arg  \min_{{\alpha,\eta}} \enspace \sum_{k=1}^{K} \left|  \hat{u}(s_k/{P})  - {u}(s_k/{P};\alpha, \eta)  \right|^{2} w_k  
 \label{eq-STSaS-CF-ln-ratio-LS} 
 \end{split} 
 \end{equation} 
 where $K$ is set to $10$ for balancing computational efficiency and fitting accuracy. 
 This weighted least squares problem is solved using two successive two-dimensional searches, as done in \cite{Gini2000-GK}. That is, a coarse grid search is first performed to narrow the search space to specific point, around which a finer search is then carried out to increase the likelihood of finding the global optimum. 
 
 Finally, using ${\hat\varphi_\mathbf{X}(\xi_\text{ref},\xi_\text{ref})}$ to approximate ${\varphi_\mathbf{X}(\xi_\text{ref},\xi_\text{ref})}$, the scale parameter $\gamma$ is estimated by 
 \begin{equation} 
 \begin{split} 
 \hat \gamma  = \frac{{{{\hat \eta }^{\hat \alpha/2 }}\ln \left[ {\frac{1}{L}\sum\nolimits_{i = 1}^L {{J_0}\left( {{s_{{\rm{ref}}}}{r_i}} \right)} } \right]}}{{1 - {{\left( {\hat \eta s_{{\rm{ref}}}^2 + 1} \right)}^{\hat \alpha /2}}}}. 
 \label{eq-estimated-gamma} 
 \end{split} 
 \end{equation}

 \begin{figure*}[t] 
 	\centering
 	\subfloat[]{ 
 		\includegraphics[width=0.32\linewidth]{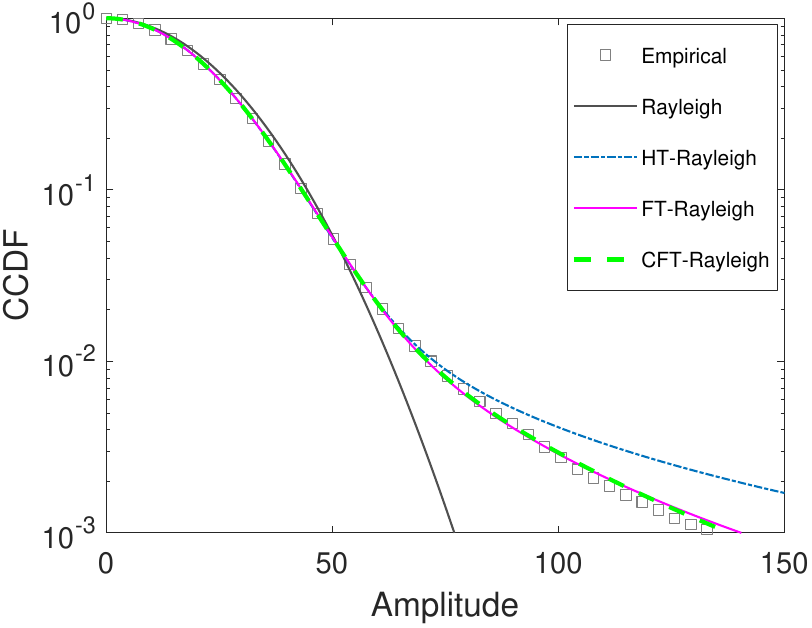}    
 	} 
 	\subfloat[]{ 
 		\includegraphics[width=0.32\linewidth]{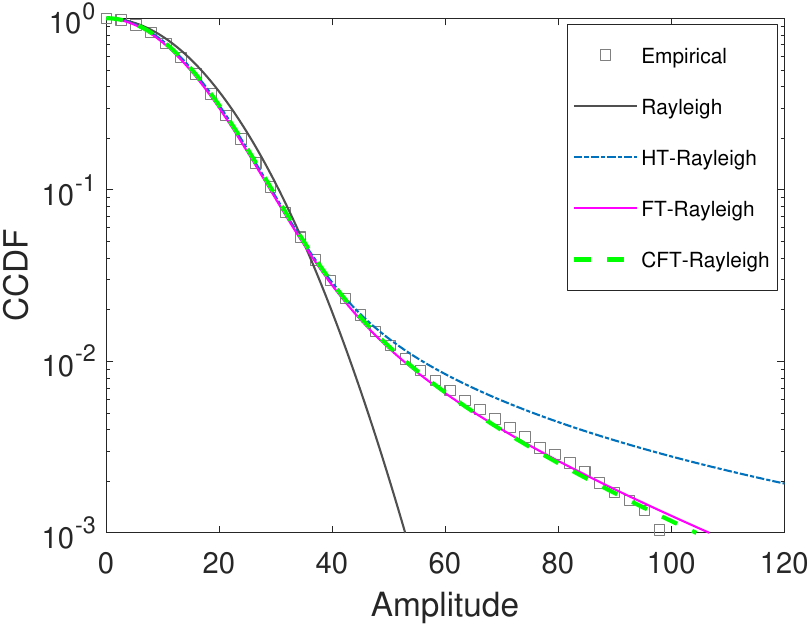}   
 	}	
 	\subfloat[]{ 
 		\includegraphics[width=0.318\linewidth]{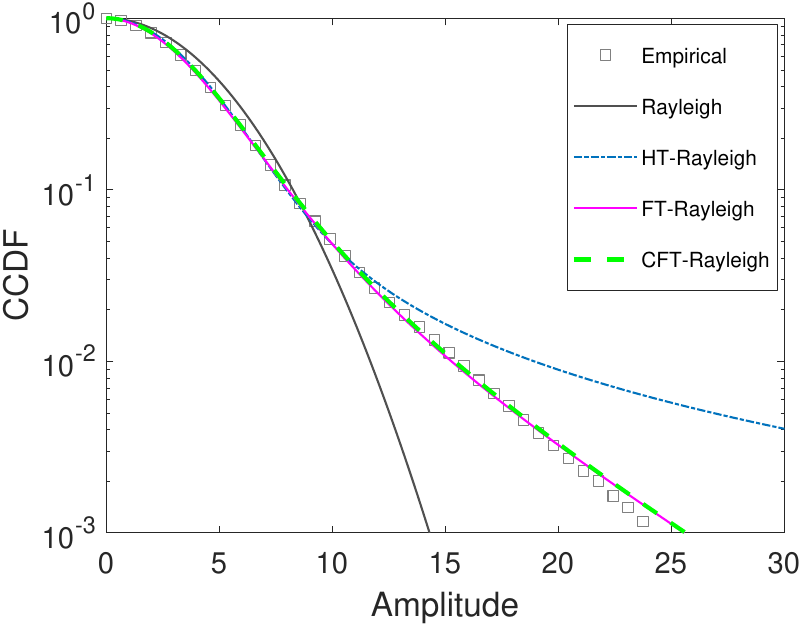}
 	}	
 	\caption{CCDF fit for McMaster IPIX radar Data\#4 of different polarizations. (a) HH polarization, range gate 16. (b) HV polarization, range gate 16. (c) VV polarization, range gate 16. }  
 	\label{fig-Data-IPIX-AMP-fit} 
 \end{figure*}

 \section{Real-World Data Analysis} 
 \label{sec:experimant}  
 This section evaluates the CFT-Rayleigh model's performance in fitting real-world sea clutter data. 
 The CFT-Rayleigh model, along with the FT-Rayleigh model introduced in \cite{Liao2024-FTRay}, rely on different tempered $\alpha$S distributions to achieve tunable tail behaviors, thereby enhancing modeling flexibility. 
 The FT-Rayleigh model has already been shown to outperform existing models in terms of modeling generality under different conditions \cite{Liao2024-FTRay}. 
 To verify the proposed CFT-Rayleigh model's generality, we compare it directly with the FT-Rayleigh model using the same data sets from \cite{Liao2024-FTRay}: ‘00\_017\_TTrFA,’ ‘TFA17\_014,’ and ‘TFC15\_038’ from the Council for Scientific and Industrial Research (CSIR) Fynmeet radar \cite{2007-Herselman-CSIR}, as well as ‘19980204\_223506\_ANTSTEP’ from the McMaster IPIX radar \cite{1998-IPIX}. These data sets were recorded at low-grazing angles, covering different sea states and polarization conditions. 
 Following the methodology in \cite{Liao2024-FTRay}, to maintain consistency across the CSIR Fynmeet radar data sets, all 49170 pulses per range gate are used for ‘00\_017\_TTrFA,’ while one pulse is selected every five for ‘TFA17\_014’ and ‘TFC15\_038’ until 49170 pulses are obtained. For the McMaster IPIX radar data set ‘19980204\_223506\_ANTSTEP,’ we analyze using all 60000 pulses. 
 The selected data are referred to as ‘Data\#1’ (00\_017\_TTrFA), ‘Data\#2’ (TFA17\_014), ‘Data\#3’ (TFC15\_038), and ‘Data\#4’ (19980204\_223506\_ANTSTEP), respectively.  
 
 \subsection{Empirical Fitting and Quantitative Analysis} 
 Figs. \ref{fig-Data-CSIR-AMP-fit}(a)-(c) show the CCDF fitting results for VV-polarized CSIR Fynmeet radar data sets: Data\#1 at range gate 7, Data\#2 at range gate 36, and Data\#3 at range gate 9, respectively.  Similarly, Figs. \ref{fig-Data-IPIX-AMP-fit}(a)-(c) present the CCDF fitting results for McMaster IPIX radar Data\#4 at range gate 16, corresponding to HH, HV, and VV polarizations, respectively. 
 In addition to the FT-Rayleigh and CFT-Rayleigh models, the Rayleigh and HT-Rayleigh models are included for comparison. Table \ref{fig-Data-fit-pars} provides the parameter estimates for these four models.

 Figs. \ref{fig-Data-CSIR-AMP-fit} and \ref{fig-Data-IPIX-AMP-fit} show that the empirical distributions exhibit lighter tails than the Rayleigh model and heavier tails than the HT-Rayleigh model. 
 As anticipated, the CFT-Rayleigh and FT-Rayleigh models significantly improve the fitting performance compared to the Rayleigh and HT-Rayleigh models. 
 Specifically, both models perform similarly, effectively adjusting the tail behavior to better match real-world clutter data in all cases. 
 
 To quantitatively assess the CCDF fitting results presented in Figs. \ref{fig-Data-CSIR-AMP-fit} and \ref{fig-Data-IPIX-AMP-fit}, Table \ref{fig-Data-fit-tests} provides the Kolmogorov-Smirnov (KS) and threshold error (TE) test results. The methods for applying the KS and TE tests are outlined in \cite{Liao2024-FTRay}, and their details are omitted here for brevity.
 As shown, the CFT-Rayleigh and FT-Rayleigh models overall exhibit much lower KS and TE statistics compared to the Rayleigh and HT-Rayleigh models, demonstrating their superior fitting performance. Notably, they pass both the KS and TE tests in all cases, highlighting their generality across different conditions.
 
 \begin{table}[t] 
 	\small 
 	\renewcommand\arraystretch{1.1} 
 	\begin{center} 
 		\caption{Parameter estimates of different models corresponding to the data shown in Figs. \ref{fig-Data-CSIR-AMP-fit} and \ref{fig-Data-IPIX-AMP-fit}} 
 		\includegraphics[width=0.96\linewidth]{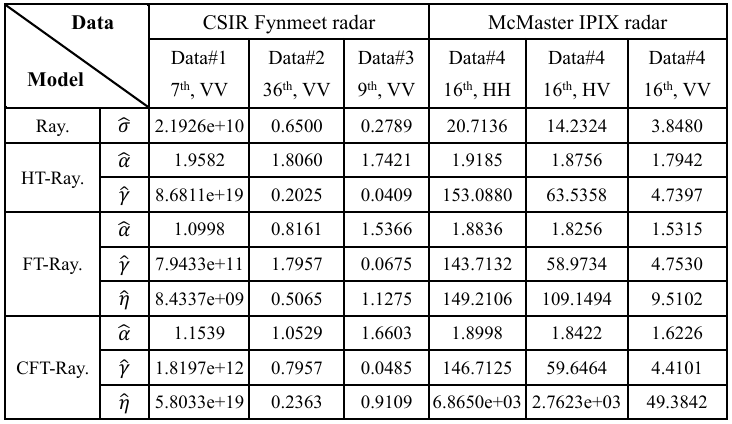} 
 		\label{fig-Data-fit-pars}
 	\end{center}
 \end{table} 
 \begin{table}[t] 
 	\small 
 	\renewcommand\arraystretch{1.1} 
 	\begin{center} 
 		\caption{The KS and TE statistics of different models corresponding to the data shown in Figs. \ref{fig-Data-CSIR-AMP-fit} and \ref{fig-Data-IPIX-AMP-fit}} 
 		\begin{threeparttable}   
 			\includegraphics[width=0.92\linewidth]{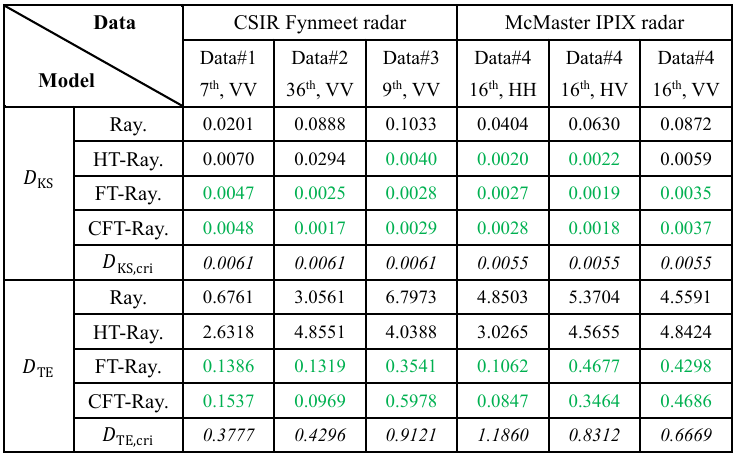}
 			\begin{tablenotes}   
 				\footnotesize              
 				\item[]Note: Cyan is used to highlight KS and TE statistics that are below their respective critical values.                      
 			\end{tablenotes} 
 		\end{threeparttable} 
 		\label{fig-Data-fit-tests} 
 	\end{center}
 \end{table}

 \subsection{Comparison of the CFT-Rayleigh and FT-Rayleigh Models}  
 The CFT-Rayleigh and FT-Rayleigh models both contain the Rayleigh and HT-Rayleigh models as special cases, with tunable tails between them. 
 As previously discussed, both models demonstrate strong performance for data with varying tail behaviors.
 However, it is important to note that they are based on different underlying assumptions when modeling complex clutter, as shown in Fig. \ref{fig-explain-Flowchart}. 
 To elaborate, the CFT-Rayleigh model utilizes the PT$\alpha$S distribution to represent the texture of the clutter, while the FT-Rayleigh model uses the ST$\alpha$S distribution to model the I/Q components of complex clutter. 
 This difference suggests that the two models are suited to different clutter types and signal processing needs. 
 
 For instance, the truncation parameter $\eta$ plays different roles: in the CFT-Rayleigh model, it controls the truncation depth for the texture, whereas in the FT-Rayleigh model, it governs the truncation depth for the I/Q components. Understanding these distinctions is key for selecting the appropriate model based on the radar environment and detection goals. If texture tail control is the focus, the CFT-Rayleigh model is likely more suitable, while the FT-Rayleigh model is better for I/Q component-based analysis. 
 
 \section{Conclusion}  
 \label{sec:conclusion} 
 In this work, we develop a more general clutter model within the CG family.  
 To achieve this, we introduce a bivariate isotropic CG-PT$\alpha$S distribution to model complex clutter and derive its correspondingly CFT-Rayleigh amplitude model. 
 We prove that the CFT-Rayleigh model is a scale mixture of Rayleighs. 
 Moreover, we provide a characteristic function-based parameter estimation method for the CFT-Rayleigh model. 
 The CFT-Rayleigh model exhibits tunable tails between Rayleigh and HT-Rayleigh, thereby theoretically providing more flexibility in clutter modeling. 
 This flexibility is validated through the analysis of real-world data under different sea states and polarizations. 
 The future research interests include two main aspects: devising clutter simulation and target detection methods based on the CFT-Rayleigh model, and providing the approximate PDF and CCDF expressions for the CFT-Rayleigh model using the continuously advancing techniques related to the tempered $\alpha$S family of distributions. 


\end{document}